\documentclass[trackchanges]{aastex701}

\usepackage{indentfirst}
\usepackage{natbib}
\begin{document}

\title{CaliciBoost: Performance-Driven Evaluation of Molecular Representations for Caco-2 Permeability Prediction\footnote{Footnotes can be added to titles}}

\author{Huong Van Le}
% \altaffiliation{Kitt Peak National Observatory}
\affiliation{Calici Co., Ltd}
\email[show]{ch.le@calici.co}  

\author{Weibin Ren} 
% \altaffiliation{Las Campanas Observatory}
\affiliation{Calici Co., Ltd}
\email{wb.ren@calici.co} 

\author{Junhong Kim}
\affiliation{Calici Co., Ltd}
\email{jh.kim@calici.co}

\author{Yukyung Yun}
\affiliation{Calici Co., Ltd}
\email{yk.yun@calici.co}

\author{Young Bin Park}
\affiliation{Calici Co., Ltd}
\email{yb.park@calici.co}

\author{Young Jun Kim}
\affiliation{Korea University}
\email{yk46@korea.ac.kr}

\author{Bok Kyung Han}
\affiliation{Korea University}
\affiliation{InsightFI Co., Ltd}
\email{hanmoo@korea.ac.kr}

\author{Inho Choi}
\affiliation{Yeungnam University}
\email{inhochoi@ynu.ac.kr}

\author{Jong IL Park}
\affiliation{Chungnam National University}
\email{jipark@cnu.ac.kr}

\author{Hwi-Yeol Yun}
\affiliation{Chungnam National University}
\email{hyyun@cnu.ac.kr}

\author{Jae-Mun Choi}
\affiliation{Calici Co., Ltd}
\affiliation{Korea University}
\affiliation{Chungnam National University}
\email{jm.choi@calici.co}

\correspondingauthor{Bok Kyung Han}
\email[show]{hanmoo@korea.ac.kr}

\correspondingauthor{Jae-Mun Choi}
\email[show]{jm.choi@calici.co}

% \author[0000-0000-0000-0003,sname=Asia,gname=Mountain]{Asia Mountain}
% \altaffiliation{Astrosat Post-Doctoral Fellow}
% \affiliation{Tata Institute of Fundamental Research, Department of Astronomy}
% \email{fakeemail5@google.com}

% \author[0000-0000-0000-0004]{Coral Australia}
% \affiliation{James Cook University, Department of Physics}
% \email{fakeemail6@google.com}

% \author[gname=IceSheet]{Penguin Antarctica}
% \affiliation{Amundsen–Scott South Pole Station}
% \email{fakeemail7@google.com}

\begin{abstract}

% Huong-Van Le †, Weibin Ren, Junhong Kim, Yukyung Yun, Sihan Lee, Young Bin Park, Inho Choi, Jong IL Park, Hwi-Yeol Yun, Young Jun Kim, Bok Kyung Han ‡, Jae-Mun Choi ‡

Caco-2 permeability serves as a critical in vitro indicator to predict oral absorption of drug candidates during early-stage drug discovery. To improve the precision and efficiency of computational predictions, we systematically investigated the impact of eight types of molecular feature representation including 2D / 3D descriptors, structural fingerprints and deep learning-based embeddings combined with automated machine learning techniques to predict Caco-2 permeability. Using two datasets of differing scale and diversity (TDC benchmark and curated OCHEM data), we assessed model performance across representations and identified PaDEL, Mordred, and RDKit descriptors as particularly effective for Caco-2 prediction. Notably, the AutoML-based model CaliciBoost achieved the best MAE performance. Furthermore, for both PaDEL and Mordred representations, the incorporation of 3D descriptors resulted in a 15.73\% reduction in MAE compared to using 2D features alone, as confirmed by feature importance analysis. These findings highlight the effectiveness of AutoML approaches in ADMET modeling and offer practical guidance for feature selection in data-limited prediction tasks.

% \textbf{Scientific contribution}
% This work provides a systematic benchmarking of eight molecular feature representation types in conjunction with AutoML for Caco-2 permeability prediction. It highlights the critical role of 3D descriptors in enhancing predictive accuracy and establishes a top PaDEL-based-feature AutoML model. The study also emphasizes the value of interpretable feature selection (via SHAP and permutation importance), offering insights into feature contributions and generalizable modeling strategies for cheminformatics applications.

\end{abstract}



\section{Introduction}
Caco-2 cell permeability is a widely used in vitro proxy for assessing the intestinal absorption of drug candidates in early-stage drug discovery. Accurately modeling this property enables effective compound prioritization, optimizes experimental resources, and reduces both the cost and time of ADMET screening. Given that oral bioavailability is a critical determinant of clinical success, predictive modeling of Caco-2 permeability plays a central role in rational drug design and high-throughput screening pipelines. However, despite its importance, building robust predictive models remains challenging due to limited data availability and the complexity of molecular feature engineering. Caco-2 permeability has long been utilized as a representative in vitro indicator for predicting the oral absorption of drug candidates in humans \cite{ref1}. Various machine learning approaches have been developed to model this property, typically relying on molecular descriptors and structural fingerprints. Early studies predominantly used random forest or support vector machines with basic physicochemical properties such as molecular weight, topological polar surface area (TPSA), LogP, hydrogen bond donors/acceptors and fingerprints like Morgan (ECFP), Avalon, and MACCS keys. Descriptor sets computed using tools such as RDKit have been widely adopted for their ability to encode structural and electronic features, while other studies such as PaDEL, Mordred, and CDDD incorporate richer 3D or learned features \cite{ref2, ref3, ref4, ref13, ref15}.

Although these studies have made progress, they typically focused on a limited subset of features or a single modeling approach. Few have conducted systematic comparisons across a wide spectrum of fingerprint and descriptor types. Moreover, the relative contributions of different representations to model performance and generalization remain poorly quantified. This gap limits our ability to make informed decisions when selecting molecular features for Caco-2 prediction tasks. Recent years have witnessed growing interest in deep learning approaches such as graph neural networks (GNNs), which can learn molecular representations directly from graph-structured input without handcrafted features. Graph Convolutional Networks (GCNs), for instance, have been explored using frameworks like DeepPurpose for ADMET-related prediction tasks, including Caco-2 permeability. However, these models require large datasets to generalize effectively. On small to medium-sized datasets such as Caco-2, their performance is often hindered by limited data availability \cite{ref5}. This is evident in benchmark results from the Therapeutics Data Commons (TDC), where classical ensemble models such as MapLight, BaseBoosting, and XGBoost consistently outperform CNN and GNN based models.

Given these limitations, selecting the optimal molecular feature representation is critical. Fingerprints such as Morgan \cite{ref6}, Avalon \cite{ref7}, ErG \cite{ref8}, and MACCS \cite{ref12} offer efficient substructure-based encodings. In parallel, descriptors derived from RDKit \cite{ref9, ref10, ref11}, PaDEL \cite{ref13}, Mordred \cite{ref3}, and CDDD \cite{ref15} capture physicochemical and structural properties, often providing complementary information. However, balancing their use, especially when datasets are not large, requires careful experimentation and validation. To address the complexity and interdependence of model development tasks, Automated Machine Learning (AutoML) has emerged as a powerful paradigm. AutoML automates key components of the machine learning pipeline, including feature selection, preprocessing, algorithm choice, and hyperparameter optimization. In cheminformatics, where datasets are typically high-dimensional, heterogeneous, and small to medium in size, AutoML offers a scalable, reproducible, and expert-free approach to model building. It is increasingly applied in QSAR modeling, ADMET property prediction, and virtual screening \cite{ref16, ref17, ref18, ref19, ref20}.

Among various frameworks, AutoGluon was selected in this study due to its superior performance on high-dimensional tabular data, strong ensemble learning capabilities, and efficient handling of missing or sparse inputs. It combines multiple model types such as LightGBM, XGBoost, CatBoost, neural networks, and k-NN and performs joint optimization using Bayesian search strategies \cite{ref21, ref22}. AutoGluon also supports preprocessing operations like normalization, categorical encoding, and imputation, and its automated ensemble construction boosts generalization while minimizing manual tuning. These characteristics make AutoGluon especially well-suited for cheminformatics tasks, where model interpretability, consistency, and scalability are essential. This study aims to fill the gap in systematic benchmarking of molecular feature representations for Caco-2 permeability prediction. We evaluate eight distinct feature types including Morgan, Avalon, ErG, MACCS fingerprints and descriptors from RDKit, PaDEL, Mordred, and CDDD across two datasets with different sizes and properties. Using AutoGluon, we assess individual predictive performance, investigating how representation choice affects accuracy, robustness, and feature importance in a data-limited setting. The significance of this study lies in its contribution to practical cheminformatics workflows. By identifying optimal molecular features and demonstrating the utility of AutoML, this work provides actionable guidance for researchers building ADMET prediction models under real-world constraints. Furthermore, the findings support evidence-based decisions in featurization and model selection, advancing reproducibility and effectiveness in drug discovery pipelines.

\section{Experimental Design}
This study employs two datasets, TDC and OCHEM, to predict Caco-2 permeability using eight different molecular feature representations: Morgan FP, Avalon FP, ErG FP, RDKit descriptors, MACCS FP, PaDEL, Mordred, and CDDD. Each feature representation is used to generate features to train in AutoML model. Feature importance is assessed via permutation importance and SHAP values, followed by selection of top-ranked features. These top features are used to retrain the model, and hyperparameters are optimized using Bayesian optimization to identify the best-performing feature representation for Caco-2 prediction. The final models are evaluated using MAE, RMSE, R², and Pearson correlation to identify the most robust approach for Caco-2 prediction. The full pipeline of the experimental design is illustrated in Figure 1.

\begin{figure*}[ht!]
\plotone{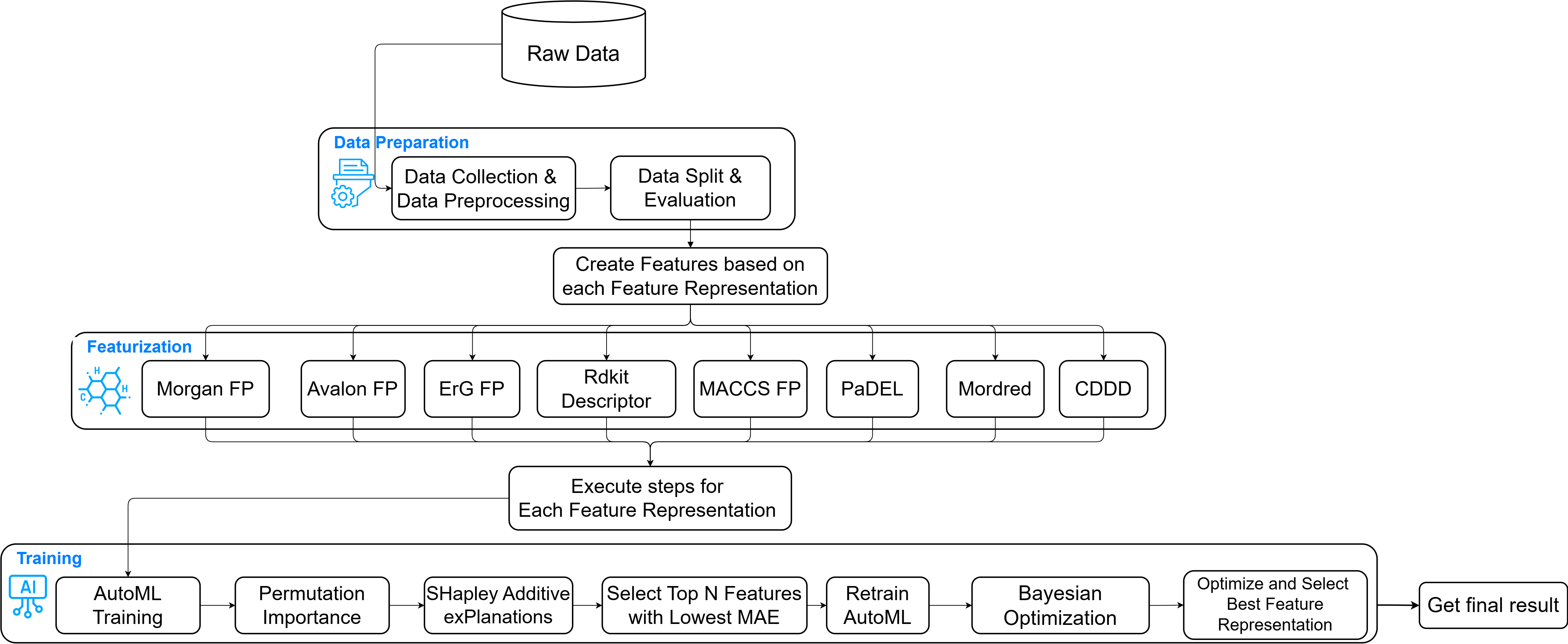}
\caption{Overall workflow for Caco-2 permeability prediction using AutoML and multi-representation molecular features %\href{https://aas.org/posts/news/2023/08/aas-open-access-publishing-model-open-transparent-and-fair}{2023 post}.
\label{fig:general}}
\end{figure*}

\subsection{Datasets}
\subsubsection{TDC}
TDC$\cdot$Caco2\_Wang dataset provided by the Therapeutics Data Commons (TDC) was used to train AutoML models for Caco-2 permeability prediction. This dataset contains 906 compounds with experimentally measured Caco-2 Papp values, curated from Wang et al. \cite {ref25}. It adopts a scaffold-based data split, commonly used to evaluate generalization to structurally novel compounds. Each entry includes a SMILES string and the corresponding permeability value, making it well-suited for QSAR modeling in standardized machine learning workflows.

\subsubsection{OCHEM}
A curated dataset of Caco-2 permeability values was obtained from the OCHEM (Online Chemical Modeling Environment) platform, a web-based system for managing and automating QSAR modeling \cite {ref26}. The dataset contains 9,402 compound entries with experimental apparent permeability (Papp) values measured across Caco-2 cell monolayers. Each entry includes SMILES, Papp values, and additional metadata such as compound name, PubMed ID, pH, temperature, and P-gp inhibition status. Most records were measured under standard assay conditions (pH 7.4, 37°C), while others lacked complete metadata or were recorded under non-standard conditions (e.g., pH 6.5), requiring preprocessing before model training.

\subsection{Data Preprocessing}
\subsubsection{TDC}
In this study, we used the scaffold split provided by TDC, which partitions the dataset into:
\begin{itemize}
    \item Training set: 728 compounds
    \item Test set: 182 compounds
\end{itemize}

\begin{figure*}[ht!]
\plotone{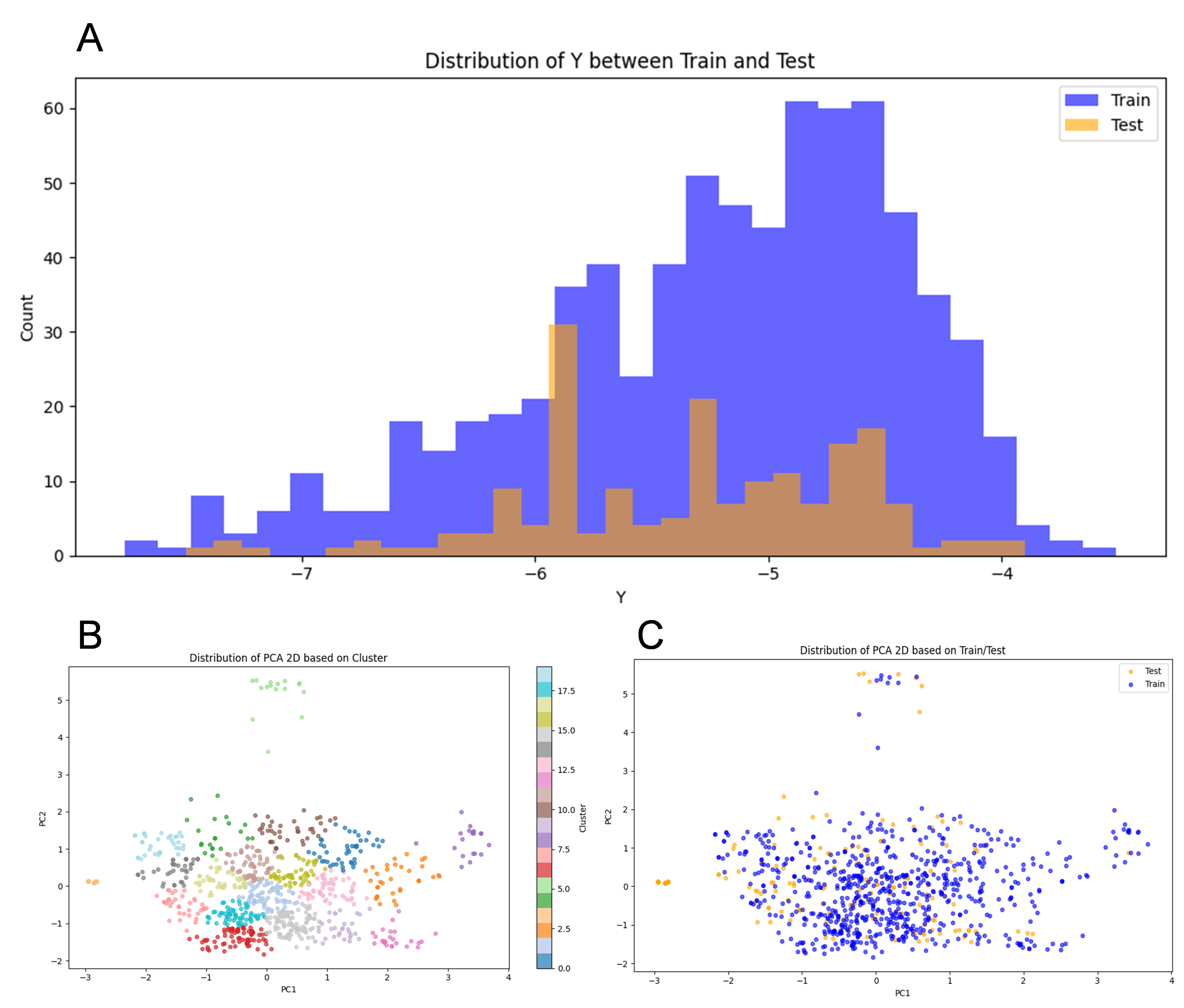}
\caption{Overview of data distribution and structural clustering in the TDC dataset. (A) Distribution of Caco-2 values in the training and test sets. (B) PCA projection of molecular structures colored by structural clusters. (C) PCA projection showing separation between train and test sets.
%\href{https://aas.org/posts/news/2023/08/aas-open-access-publishing-model-open-transparent-and-fair}{2023 post}.
\label{fig:general}}
\end{figure*}

Although the TDC dataset provides predefined train and test splits, we conducted an additional examination of the data distribution and structural clustering to verify the integrity of the split. As visualized in Figure 2, the distribution of Caco-2 permeability values (Y) remains consistent across the training and test subsets. Furthermore, Principal Component Analysis (PCA) and clustering reveal that chemical structures are well spread and that the test set adequately covers the structural diversity observed in the training set. These observations affirm the reliability and representativeness of the TDC split for robust model evaluation.

\subsubsection{OCHEM}
To ensure consistency and experimental relevance, several preprocessing steps were applied to the OCHEM Caco-2 permeability dataset prior to model training. First, only entries with a permeability direction of apical-to-basolateral (A→B) were retained, as this direction aligns with standard practice in modeling intestinal absorption. The entries explicitly labeled as basolateral-to-apical (B→A) were excluded, while entries with missing direction metadata were assumed to be A→B, following common assumptions in permeability assays.
Next, filtering was applied based on pH conditions. Since most Caco-2 experiments are conducted at pH 7.4, entries reporting pH values different from 7.4 were removed to avoid introducing variability. In total, 573 entries were excluded in this step. Records with missing pH values were imputed as pH 7.4, assuming standard assay conditions.
In addition, temperature metadata were considered to maintain consistency with typical biological assays conducted at 37°C. Entries with reported temperatures differing from 37°C or missing temperature information were removed to ensure uniform experimental conditions.

Finally, to stabilize the regression task and reduce the skewness in permeability values, all Papp measurements were transformed using base-10 logarithm ($\log{10}$). This transformation is commonly used in QSAR modeling to normalize data distributions and improve model performance. After applying all preprocessing steps, a total of 5,481 curated records remained for model development.

Following data curation, a total of 5,481 high-quality entries were retained from the OCHEM dataset for further modeling. To enable cluster-aware and value-balanced splitting between training and testing subsets, the following steps were applied: 
Molecular structures were encoded using 1024-bit Morgan fingerprints, and then projected into a lower-dimensional space using Principal Component Analysis (PCA) for visualization and structural diversity assessment. In this 2D PCA space, KMeans clustering was performed to group molecules by structural similarity. In parallel, the log-transformed Caco-2 Papp values were binned into discrete intervals to allow for stratified sampling by permeability levels.

Using both cluster assignments and Papp bins, we performed a stratified split to construct training and test sets that preserved the overall distribution of both chemical structure and permeability. As a result, the dataset was divided into:
\begin{itemize}
    \item Training set: 4,377 compounds
    \item Test set: 1,095 compounds
\end{itemize}

As shown in Figure 3, the distribution of Caco-2 values between the train and test sets remains consistent. The PCA projections further confirm structural diversity, with distinct cluster separation and a balanced distribution of train and test compounds in chemical space. These results indicate that the sampling process preserved both label and structural diversity without introducing significant bias. 
\begin{figure*}[ht!]
\plotone{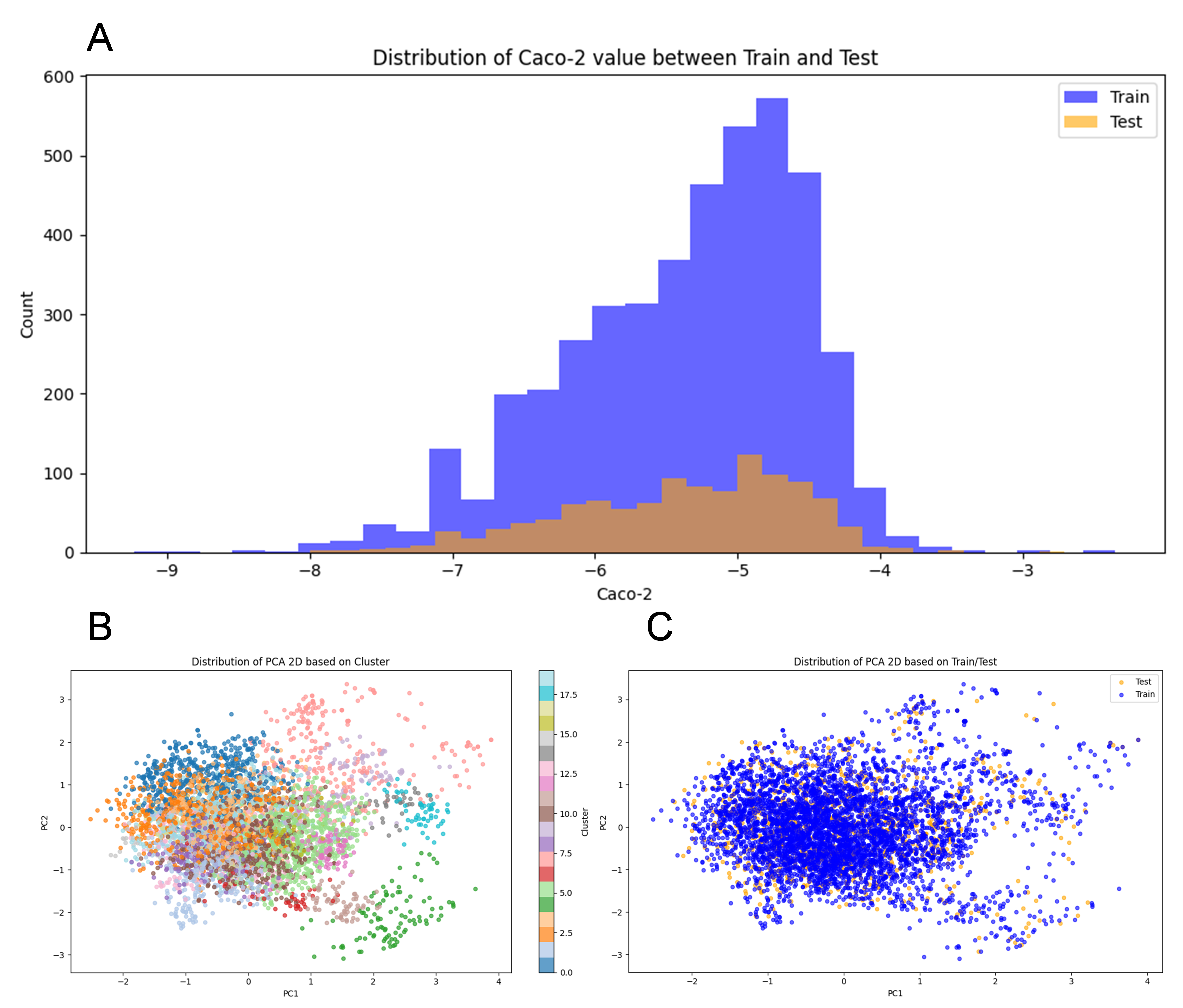}
\caption{Overview of data distribution and structural clustering in the OCHEM dataset. (A) Distribution of Caco-2 values in the training and test sets. (B) PCA projection of molecular structures colored by structural clusters. (C) PCA projection showing separation between train and test sets.
%\href{https://aas.org/posts/news/2023/08/aas-open-access-publishing-model-open-transparent-and-fair}{2023 post}.
\label{fig:general}}
\end{figure*}

\subsubsection{Feature Extraction}
To represent molecular structures in a machine-readable format, we extracted features using multiple cheminformatics tools and packages. Morgan, Avalon, ErG fingerprints, MACCS fingerprints, and RDKit descriptors were computed using the RDKit toolkit (ver. 2023.9.6). CDDD descriptors were generated via a pretrained sequence-to-sequence autoencoder, while PaDEL and Mordred descriptors were calculated using the padelpy (ver. 0.1.14) and Mordred Python package (community ver. 2.0.6) respectively.

\begin{itemize}
    \item Morgan fingerprints (1024 bits) were computed using GetHashedMorganFingerprint, which encodes circular atom environments based on atomic neighborhoods.
    \item Avalon fingerprints (1024 bits) were generated using GetAvalonCountFP, capturing predefined substructure patterns.
    \item ErG fingerprints (315 dimensions) were calculated via GetErGFingerprint, encoding topological relationships between pharmacophoric features.
    \item RDKit descriptors were extracted using MolecularDescriptorCalculator, yielding over 200 physicochemical and topological properties.
    \item MACCS fingerprints (167 bits) were generated using GenMACCSKeys, capturing the presence or absence of predefined structural keys commonly associated with bioactivity and chemical functionality.
    All RDKit-based features were converted to NumPy arrays and used as input for downstream machine learning models.
    \item PaDEL descriptors were calculated using the padelpy wrapper for the PaDEL-Descriptor software, producing 1,875 2D, and 3D descriptors per molecule. 
    \item Mordred descriptors were computed with the Mordred Python package, which supports over 1,800 descriptors including physicochemical, topological, and 3D properties. 
    \item CDDD descriptors were generated using a pretrained sequence-to-sequence autoencoder that translates between different SMILES representations. The encoder produces a 512-dimensional continuous embedding per molecule. We used the official CDDD Python package provided by Winter et al. (2019).
\end{itemize}

\subsubsection{AutoML Training and Optimization}
To develop predictive models for Caco-2 permeability, we employed AutoGluon-Tabular (v0.7.0), an AutoML framework optimized for tabular datasets with high-dimensional feature spaces. For each feature representation type - Morgan, Avalon, ErG, RDKit descriptors, MACCS, CDDD, PaDEL, and Mordred - a separate AutoGluon model was trained using the best\_quality preset. The target variable was Caco-2 permeability, and the primary evaluation metric was mean absolute error (MAE), consistent with the Therapeutics Data Commons (TDC) leaderboard. Additional metrics including root mean squared error (RMSE), R-squared (R²), and Pearson correlation coefficient (r) were also calculated to provide a more comprehensive performance assessment.
Model performance across feature representation types was compared based on MAE. For each feature representation type, the top features were selected for further analysis, where we identified top-ranked features in each set using permutation importance and SHAP values. Permutation feature importance is a model-agnostic technique that estimates the importance of each feature by measuring the decrease in model performance when the feature’s values are randomly shuffled, thereby breaking its relationship with the target variable \cite {ref27}. SHAP (SHapley Additive exPlanations) is a unified framework for interpreting model predictions by assigning each feature an importance value for individual predictions, combining principles from cooperative game theory with additive feature attribution methods to ensure consistency and local accuracy \cite {ref28}. The model was then retrained using only the most informative features of each feature representation type to assess the impact of dimensionality reduction on performance and interpretability. In the final stage, we applied Bayesian optimization to fine-tune the model's hyperparameters. This procedure aimed to maximize predictive accuracy while maintaining model robustness. 

\section{Results and Discussion} \label{sec:style}
\subsection{Effect of Dataset Split - TDC vs. OCHEM} \label{subsec:style}

To evaluate the influence of dataset split strategy on model performance, we systematically compared model outcomes using the same feature representations and AutoML frameworks under both the TDC scaffold split and the OCHEM custom split. The results are summarized in Table 1 (TDC) and Table 2 (OCHEM), with a graphical comparison provided in Figure 4 (TDC) and Figure 5 (OCHEM).

\begin{table*}[ht]
\centering
\caption{Performance comparison of models trained with the TDC dataset across different feature representation types and subsets}
\label{tab:results}
\begin{tabular}{|c|c|c|c|c|c|c|c|c|}
\hline
\textbf{Exp. No.} & \textbf{Feature Type} & \textbf{Feature Subset} & \textbf{Dataset} & \textbf{Best Model} & \textbf{MAE} & \textbf{RMSE} & \textbf{R$^2$} & \textbf{Pearson R} \\
\hline
1  & Morgan FP        & All                & TDC      & LightGBM         & 0.3007 & 0.3729 & 0.7046 & 0.8410 \\
2  & Morgan FP        & Top                & TDC      & WeightedEnsemble & 0.2880 & 0.3620 & 0.7220 & 0.8510 \\
3  & Avalon FP        & All                & TDC      & WeightedEnsemble & 0.2941 & 0.3684 & 0.7118 & 0.8437 \\
4  & Avalon FP        & Top                & TDC      & NeuralNetTorchNN & 0.2750 & 0.3450 & 0.7480 & 0.8740 \\
5  & ERG FP           & All                & TDC      & CatBoost         & 0.3172 & 0.3914 & 0.6747 & 0.8226 \\
6  & ERG FP           & Top                & TDC      & LightGBM         & 0.3080 & 0.3860 & 0.6840 & 0.8310 \\
7  & RDKit Descriptor & All                & TDC      & LightGBMLarg     & 0.2741 & 0.3478 & 0.7432 & 0.8640 \\
8  & RDKit Descriptor & Top                & TDC      & WeightedEnsemble & 0.2670 & 0.3390 & 0.7560 & 0.8700 \\
9  & MACCS FP         & All                & TDC      & LightGBMLarge    & 0.3165 & 0.3952 & 0.6683 & 0.8200 \\
10 & MACCS FP         & Top                & TDC      & WeightedEnsemble & 0.2870 & 0.3770 & 0.6980 & 0.8360 \\
11 & PaDEL            & All                & TDC      & Gradient Boosting& 0.3058 & 0.3826 & 0.6887 & 0.8331 \\
12 & PaDEL            & 2D                 & TDC      & Gradient Boosting& 0.3037 & 0.3785 & 0.6953 & 0.8363 \\
13 & PaDEL            & 3D                 & TDC      & Gradient Boosting& 0.4277 & 0.5663 & 0.3177 & 0.5714 \\
14 & PaDEL            & Top                & TDC      & XGBoosting       & 0.2560 & 0.3224 & 0.7788 & 0.8839 \\
15 & Mordred          & All                & TDC      & Gradient Boosting& 0.3033 & 0.3835 & 0.6886 & 0.8306 \\
16 & Mordred          & 2D                 & TDC      & Gradient Boosting& 0.2916 & 0.3649 & 0.7181 & 0.8500 \\
17 & Mordred          & 3D                 & TDC      & Random Forest    & 0.3883 & 0.4775 & 0.5172 & 0.7352 \\
18 & Mordred          & Top                & TDC      & Gradient Boosting& 0.2613 & 0.3413 & 0.7533 & 0.8687 \\
19 & CDDD             & All                & TDC      & LightGBMXT       & 0.3565 & 0.4716 & 0.5278 & 0.7308 \\
20 & CDDD             & Top                & TDC      & CatBoost         & 0.3590 & 0.4750 & 0.5210 & 0.7290 \\
\hline
\end{tabular}
\end{table*}

Across all eight feature representation types - Morgan fingerprints, Avalon fingerprints, ErG fingerprints, RDKit descriptors, MACCS fingerprints, PaDEL, Mordred, and CDDD - models trained and evaluated on the TDC split consistently achieved lower MAE, RMSE, and higher R² and Pearson correlation, indicating better predictive accuracy. This trend highlights TDC’s stability as a benchmark for evaluating Caco-2 permeability models.

\begin{table*}[ht]
\centering
\caption{Performance comparison of models trained with the OCHEM dataset across different feature representation types and subsets}
\label{tab:results}
\begin{tabular}{|c|c|c|c|c|c|c|c|c|}
\hline
\textbf{Exp. No.} & \textbf{Feature Type} & \textbf{Feature Subset} & \textbf{Dataset} & \textbf{Best Model} & \textbf{MAE} & \textbf{RMSE} & \textbf{R$^2$} & \textbf{Pearson R} \\
\hline
1  & Morgan FP        & All                & OCHEM    & NeuralNetFastAI  & 0.3592 & 0.5030 & 0.6030 & 0.7773 \\
2  & Morgan FP        & Top                & OCHEM    & WeightedEnsemble & 0.3860 & 0.5320 & 0.5570 & 0.7480 \\
3  & Avalon FP        & All                & OCHEM    & WeightedEnsemble & 0.3558 & 0.4949 & 0.6156 & 0.7851 \\
4  & Avalon FP        & Top                & OCHEM    & CatBoost         & 0.3730 & 0.5150 & 0.5840 & 0.7650 \\
5  & ERG FP           & All                & OCHEM    & WeightedEnsemble & 0.3824 & 0.5232 & 0.5704 & 0.7559 \\
6  & ERG FP           & Top                & OCHEM    & XGBoost          & 0.3700 & 0.5100 & 0.5910 & 0.7690 \\
7  & RDKit Descriptor & All                & OCHEM    & XGBoost          & 0.3697 & 0.5139 & 0.5856 & 0.7669 \\
8  & RDKit Descriptor & Top                & OCHEM    & WeightedEnsemble & 0.3530 & 0.4970 & 0.6120 & 0.7840 \\
9  & MACCS FP         & All                & OCHEM    & LightGBMLarge    & 0.3752 & 0.5274 & 0.5634 & 0.7506 \\
10 & MACCS FP         & Top                & OCHEM    & WeightedEnsemble & 0.3790 & 0.5320 & 0.5570 & 0.7460 \\
11 & PaDEL            & All                & OCHEM    & XGBoosting       & 0.3870 & 0.5231 & 0.5716 & 0.7569 \\
12 & PaDEL            & 2D                 & OCHEM    & Random Forest    & 0.3944 & 0.5266 & 0.5659 & 0.7579 \\
13 & PaDEL            & 3D                 & OCHEM    & XGBoosting       & 0.5264 & 0.6734 & 0.2902 & 0.5402 \\
14 & PaDEL            & Top                & OCHEM    & Random Forest    & 0.3826 & 0.5179 & 0.5802 & 0.7649 \\
15 & Mordred          & All                & OCHEM    & XGBoosting       & 0.4042 & 0.5463 & 0.5335 & 0.7306 \\
16 & Mordred          & 2D                 & OCHEM    & XGBoosting       & 0.3987 & 0.5423 & 0.5404 & 0.7359 \\

17 & Mordred          & 3D                 & OCHEM    & Random Forest    & 0.5008 & 0.6462 & 0.3472 & 0.5898 \\
18 & Mordred          & Top                & OCHEM    & XGBoosting       & 0.4046 & 0.5473 & 0.5317 & 0.7295 \\
19 & CDDD             & All                & OCHEM    & RandomForest     & 0.4103 & 0.5526 & 0.5207 & 0.7218 \\
20 & CDDD             & Top                & OCHEM    & WeightedEnsemble & 0.4030 & 0.5560 & 0.5150 & 0.7200 \\
\hline
\end{tabular}
\end{table*}

However, it is important to note that the OCHEM dataset, being approximately five times larger than TDC, provides a broader and more diverse chemical space. Despite this increased complexity, model performance on OCHEM remains reasonably strong across all representations, as shown in Figure 9. The results suggest that the AutoML framework retains predictive capacity even under more challenging, real-world-like conditions, demonstrating the robustness of the trained models beyond controlled benchmark splits.

\begin{figure}[ht!]
\centering
\includegraphics[width=0.9\textwidth]{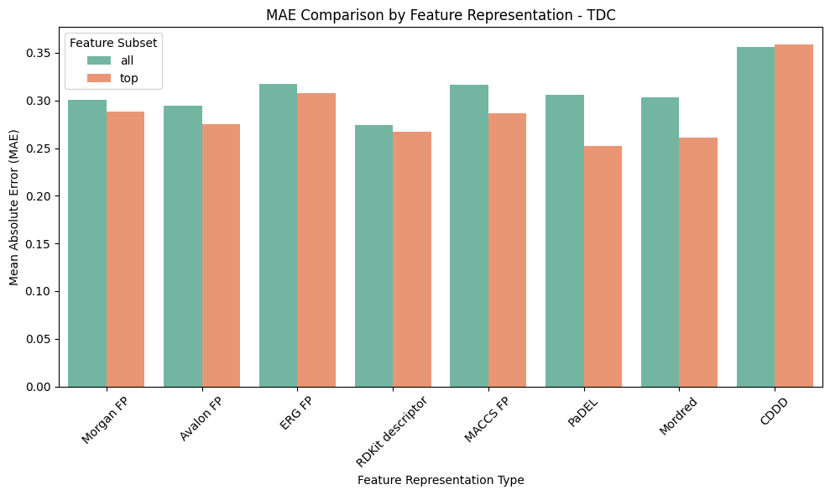}
\caption{MAE of models trained with the TDC dataset across different feature representations using both all and top features}
\label{fig:my_table}
\end{figure}

In Figure 4, we visualize MAE comparisons for both all and top features across each molecular representation on the TDC dataset. A similar analysis for the OCHEM dataset is shown in Figure 5, where overall MAE values tend to be higher. These plots further emphasize the increased difficulty of the OCHEM split, likely stemming from its greater structural diversity, larger chemical space, and less uniform data distribution, in contrast to the more stable scaffold-based split used in TDC.

\begin{figure}[ht!]
\centering
\includegraphics[width=0.9\textwidth]{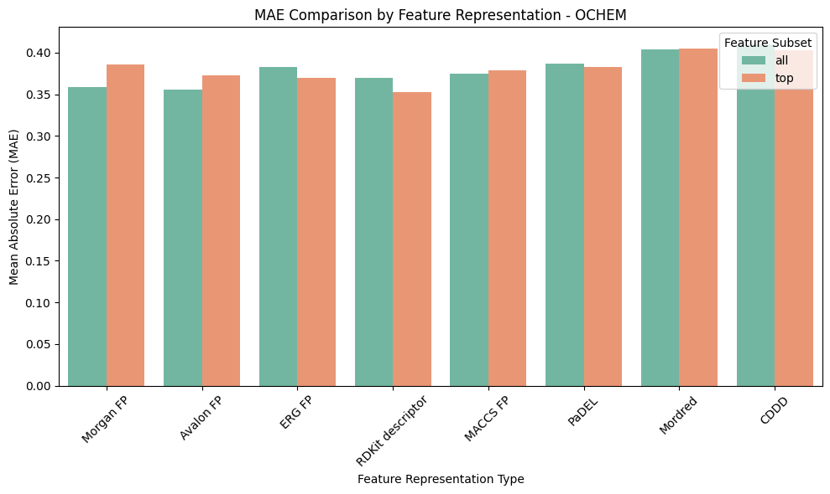}
\caption{MAE of models trained with the OCHEM dataset across different feature representations using both all and top features}
\label{fig:my_table}
\end{figure}

In summary, the TDC benchmark provides a more stable, balanced, and reproducible environment for evaluating model performance, making it well-suited for internal validation. On the other hand, the OCHEM dataset introduces a more realistic and challenging setting, thereby serving as a more rigorous testbed for assessing model robustness and generalization. These insights are critical when selecting appropriate benchmarks for model development and deployment in cheminformatics and virtual screening applications.

\subsection{Performance of Individual Feature Representations} \label{subsec:style}
\subsubsection{Using all features in each feature representation} \label{subsubsec:style}

Each molecular representation - including Morgan, Avalon, ErG, RDKit descriptors, MACCS, PaDEL, Mordred, and CDDD - was independently evaluated by training auto ML models using its entire set of features, without prior filtering or feature selection. The evaluation was conducted separately on the TDC and OCHEM splits to examine the robustness and generalizability of each representation.
On the TDC dataset, models trained on RDKit descriptors, Avalon fingerprints, and Morgan fingerprints exhibited the best performance (e.g., 0.2741, 0.2941, and 0.3007 respectively), closely following by Mordred and PaDEL (0.3058 and 0.3033), achieving the lowest MAE values and high R² (up to 0.7432) and Pearson correlation coefficients (up to 0.8640), as shown in Table 1. This indicates that these representations capture structural patterns highly relevant to Caco-2 permeability and are effective even without additional dimensionality reduction.

In contrast, on the OCHEM dataset, overall performance is not as good as TDC across all representations (see Table 2), with higher MAE and lower R², consistent with the broader observation that TDC provides a more stable benchmark (see Section 3.1: Effect of Dataset Split - TDC vs. OCHEM). However, among the representations, RDKit descriptors, Avalon fingerprints, and Morgan fingerprints still maintained relatively better performance on OCHEM.

These findings highlight that RDKit descriptors, Avalon fingerprints, Morgan fingerprints, PaDEL, Mordred - even without prior feature selection - contribute substantially to model performance in Caco-2 permeability prediction, underscoring their intrinsic value as robust input representations in AutoML-based modeling frameworks.

\subsubsection{Using top features in each feature representation } \label{subsubsec:style}
To improve both interpretability and model efficiency, we applied feature importance analysis using permutation importance and SHAP (SHapley Additive exPlanations) values for each trained model based on each representation type. This allowed us to identify the most influential features within each representation type.
Initially, we evaluated retraining models using a fixed number of top features (e.g., top 100, 50, or 30). However, this strategy did not consistently improve performance and in some cases, resulted in worse metrics compared to models trained on all features in each feature representation. This outcome suggests that the optimal number of top features is not universal across representations and that rigid cutoffs may exclude features that contribute to non-linear interactions important for prediction.
To address this, we conducted a screening experiment, evaluating model performance using the top N features, where N ranged from 1 to 100 or 1 to 200, depending on the dimensionality of each representation. This approach allowed us to identify the optimal number of informative features required to match or exceed the performance of the full model. Performance trends for each representation (MAE vs. number of top features) are provided in Figure 6 (for TDC dataset) and Figure 7 (for OCHEM dataset). This screening process allowed us to select the optimal subset of features that retained or improved predictive performance compared to using the full feature set.

\begin{figure}[ht!]
\centering
\includegraphics[width=0.9\textwidth]{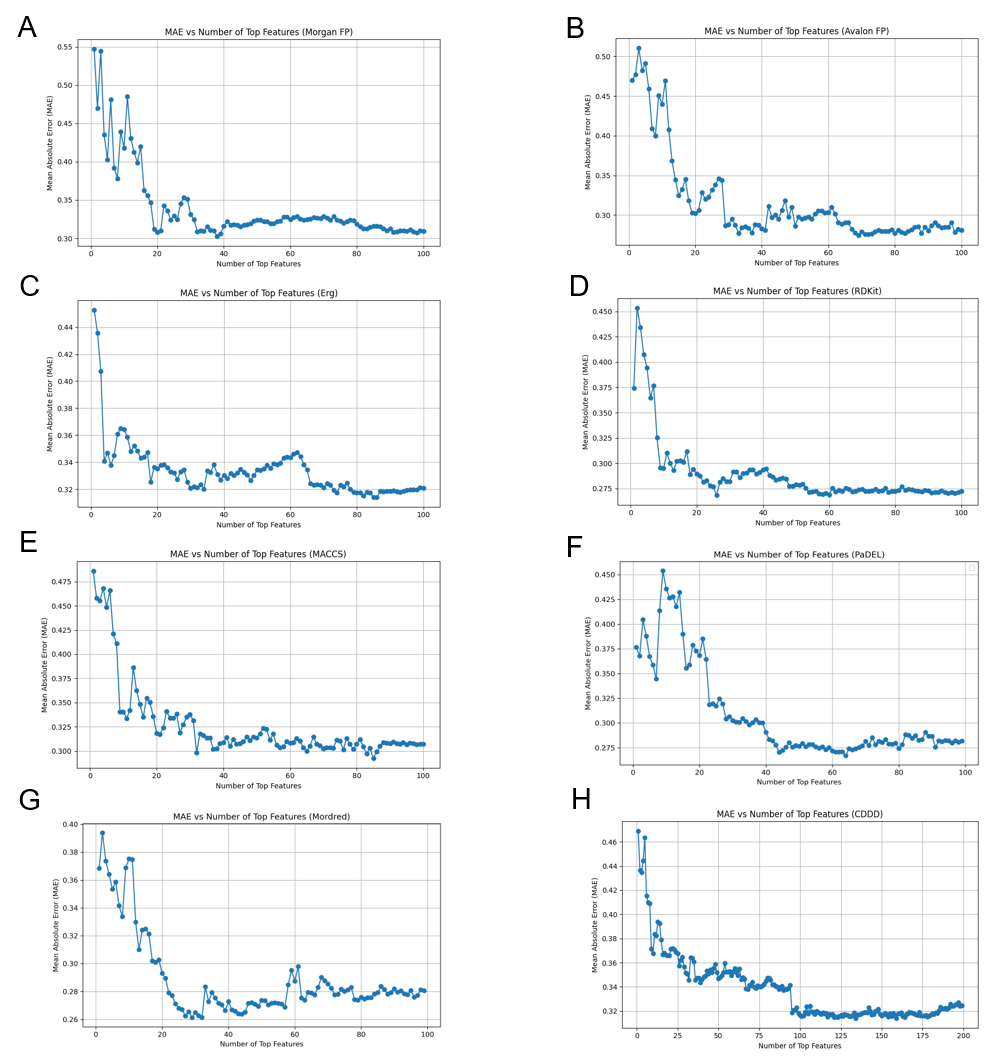}
\caption{Effect of the number of top features on model MAE across eight molecular representations on the TDC dataset. (A) Morgan fingerprints (B)Avalon fingerprints (C) ErG fingerprints (D) RDKit descriptors (E) MACCS fingerprints (F) PaDEL descriptors (G) Mordred descriptors (H) CDDD embeddings}
\label{fig:my_table}
\end{figure}

The refined feature subsets of each feature representation type were then used to retrain the models with Bayesian optimization, resulting in consistent improvements in MAE and correlation metrics. This analysis not only improved model performance but also provided insight into which chemical substructures or properties were most predictive of Caco-2 permeability within each molecular representation.

\begin{figure}[ht!]
\centering
\includegraphics[width=0.9\textwidth]{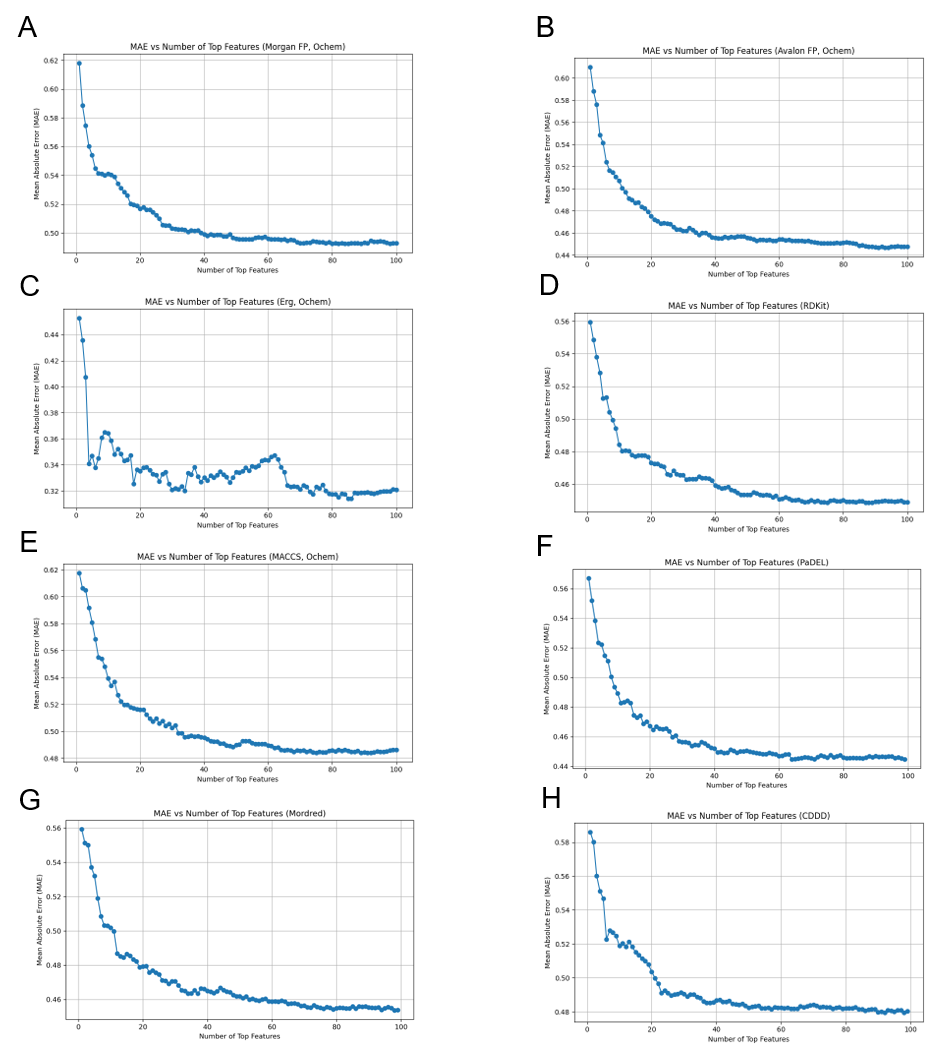}
\caption{Effect of the number of top features on model MAE across eight molecular representations on the OCHEM dataset. (A) Morgan fingerprints (B)Avalon fingerprints (C) ErG fingerprints (D) RDKit descriptors (E) MACCS fingerprints (F) PaDEL descriptors (G) Mordred descriptors (H) CDDD embeddings}
\label{fig:my_table}
\end{figure}

Notably, our AutoML-based model CaliciBoost, trained on the top PaDEL features from the TDC dataset, achieved the best overall performance, with an average MAE of 0.2560, RMSE of 0.3224, R² of 0.7788, and a Pearson correlation of 0.8839, outperforming all other models evaluated in this study. The Mordred top feature model followed closely (MAE = 0.2613, RMSE = 0.3413, R² = 0.7533, Pearson r = 0.8687). The model trained with RDKit top descriptors also showed significant performance gains, with metrics of MAE = 0.2670, RMSE = 0.3390, R² = 0.7560, and Pearson r = 0.8700 (Figure 8).

\begin{figure}[ht!]
\centering
\includegraphics[width=0.9\textwidth]{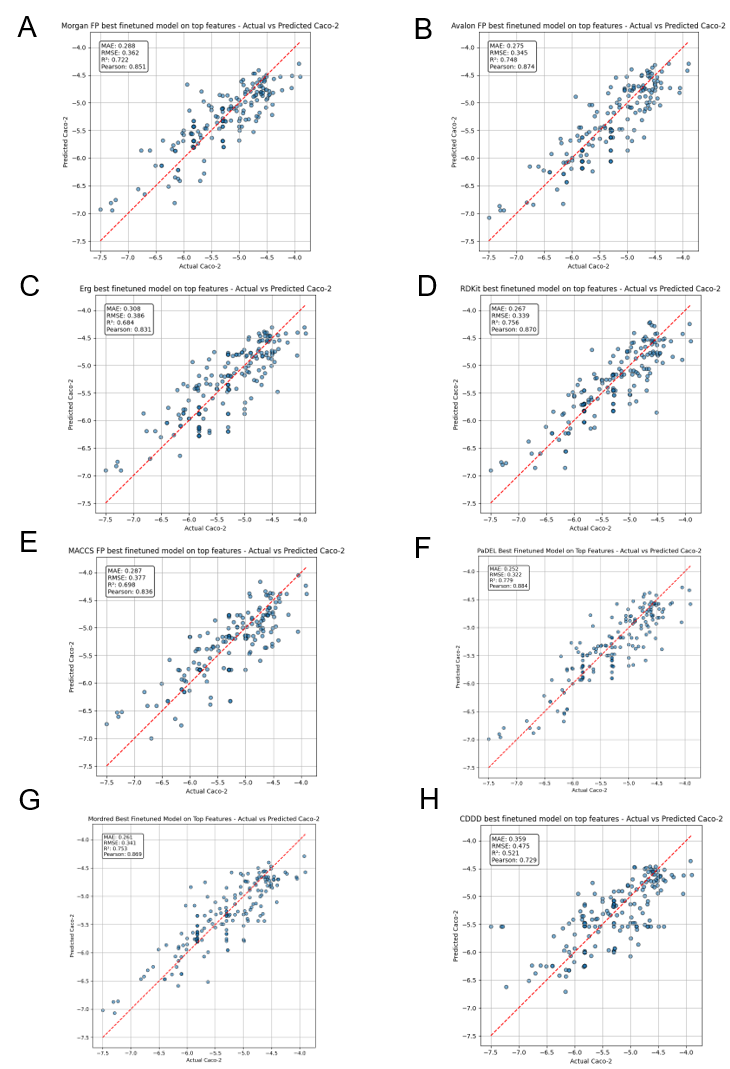}
\caption{Model performance across eight molecular feature representations on the TDC dataset. (A) Morgan fingerprints (B)Avalon fingerprints (C) ErG fingerprints (D) RDKit descriptors (E) MACCS fingerprints (F) PaDEL descriptors (G) Mordred descriptors (H) CDDD embeddings}
\label{fig:my_table}
\end{figure}

A similar trend was observed with the OCHEM dataset, where retraining models on top-ranked features followed by Bayesian optimization also led to noticeable improvements in prediction performance (Figure 9). While overall scores on OCHEM remained slightly lower due to its increased complexity and chemical diversity, the relative gains reinforce the effectiveness of this two-step refinement strategy for enhancing model quality across datasets.

\begin{figure}[ht!]
\centering
\includegraphics[width=0.9\textwidth]{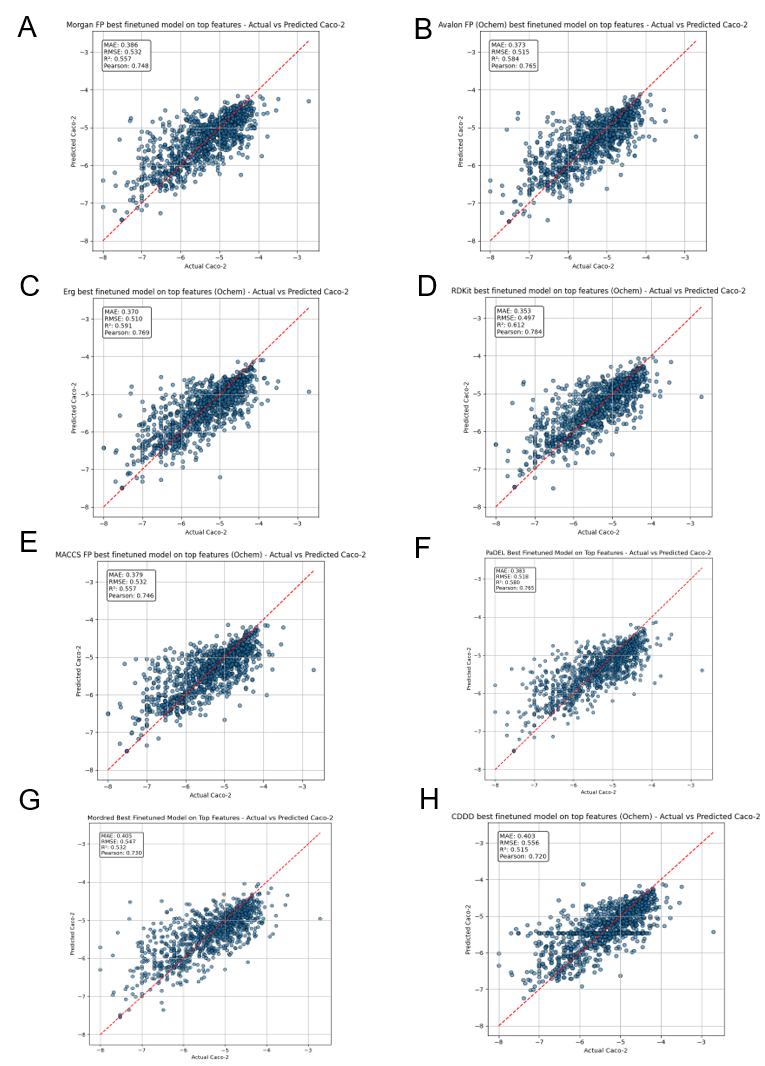}
\caption{Model performance across eight molecular feature representations on the OCHEM dataset. (A) Morgan fingerprints (B) Avalon fingerprints (C) ErG fingerprints (D) RDKit descriptors (E) MACCS fingerprints (F) PaDEL descriptors (G) Mordred descriptors (H) CDDD embeddings}
\label{fig:my_table}
\end{figure}

Among the eight molecular feature representation types evaluated in this study, only PaDEL and Mordred offer the capability to generate both 2D and 3D molecular descriptors. This provides a unique opportunity to investigate the relative importance and contribution of 3D structural information in predicting Caco-2 permeability. To this end, we conducted additional experiments to assess model performance when trained separately on 2D descriptors, 3D descriptors, and the full descriptor sets (2D + 3D) from both PaDEL and Mordred. Through this analysis, we aimed to understand whether incorporating 3D descriptors enhances model accuracy, and whether these features appear among the most informative descriptors selected during model refinement. Our results indicate that models trained with only 2D descriptors consistently underperformed compared to those utilizing the full descriptor sets (2D + 3D), particularly in terms of MAE and R² scores for both TDC (Figure 10) and OCHEM dataset (Figure 11). This suggests that 3D structural information plays a significant role in accurately modeling Caco-2 permeability.

\begin{figure}[h]
\centering
\includegraphics[width=0.9\textwidth]{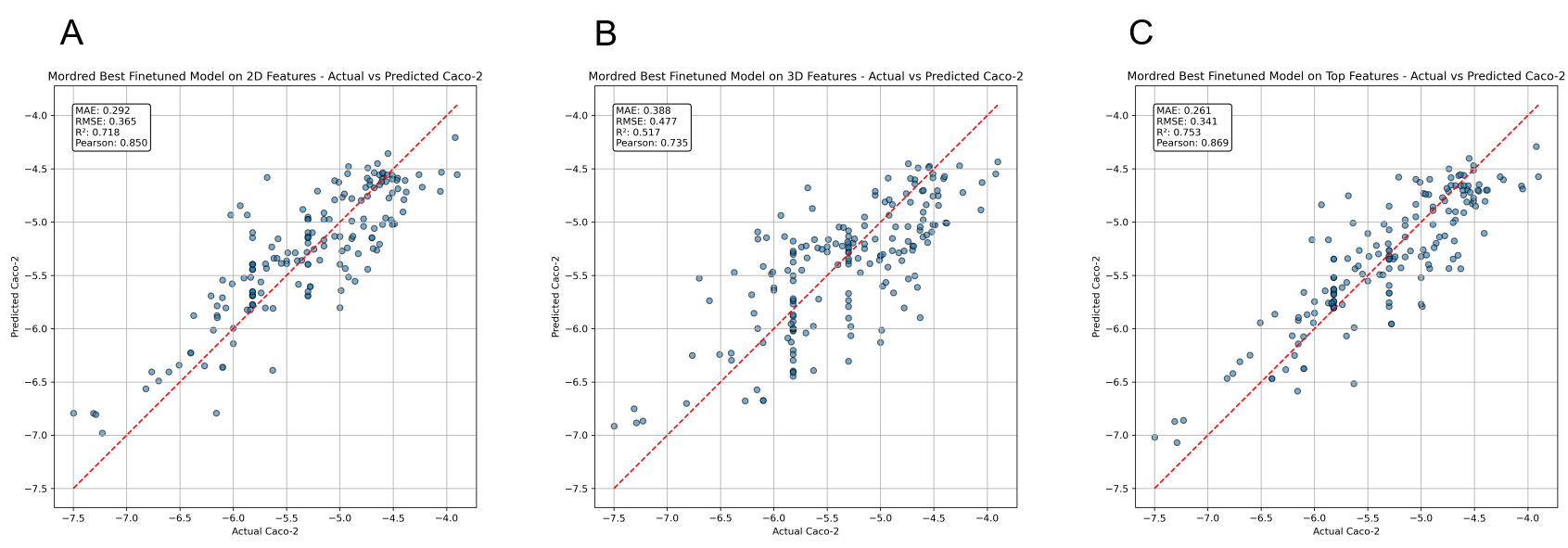}
\caption{Comparative model performance using Mordred descriptors on the TDC dataset.
(A) Model trained with only 2D descriptors. (B) Model trained with only 3D descriptors.
(C) Model trained with both 2D and 3D descriptors using top feature selection and Bayesian optimization.
}
\label{fig:my_table}
\end{figure}

Notably, when top features were selected using SHAP values and permutation importance for retraining and optimization, the selected subsets consistently included both 2D and 3D descriptors. This reinforces the conclusion that while 2D features capture essential molecular substructures, 3D descriptors contribute complementary spatial and geometric information crucial for permeability prediction. A complete list of the selected top features, along with their corresponding SHAP values, is provided in Supplementary Information 1 for both PaDEL and Mordred representations.

\begin{figure}[h]
\centering
\includegraphics[width=0.9\textwidth]{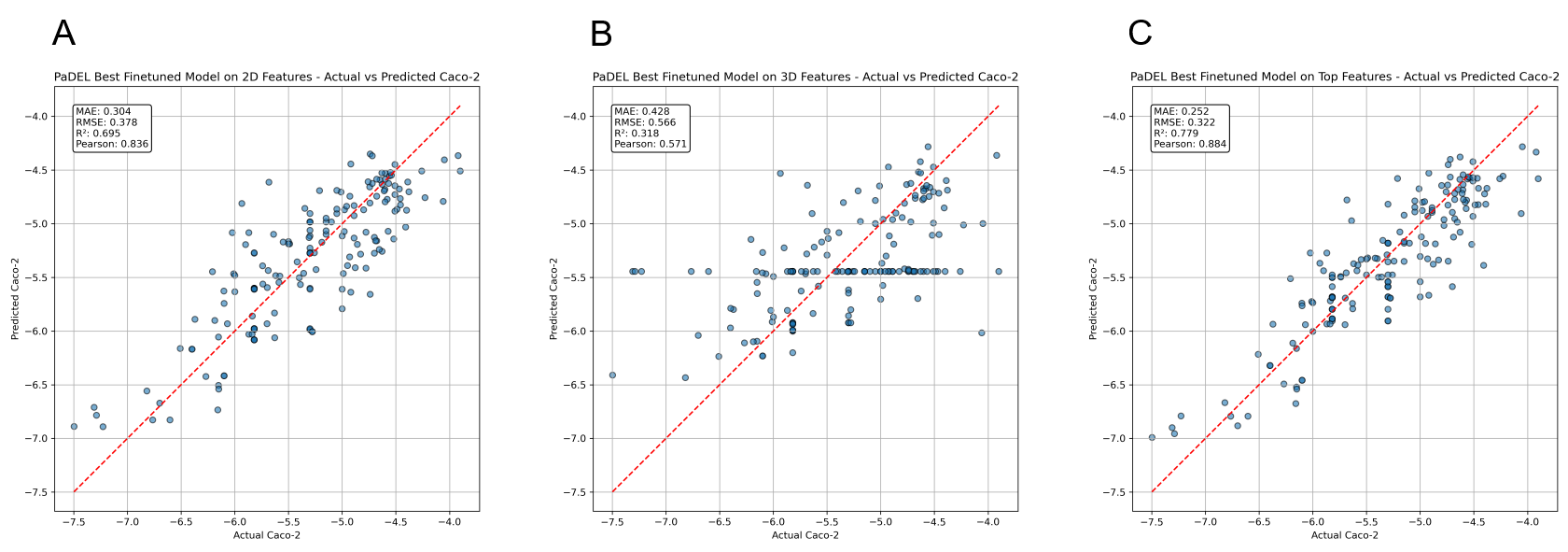}
\caption{Comparative model performance using Mordred descriptors on the OCHEM dataset.
(A) Model trained with only 2D descriptors. (B) Model trained with only 3D descriptors.
(C) Model trained with both 2D and 3D descriptors using top feature selection and Bayesian optimization.
}
\label{fig:my_table}
\end{figure}

\section{Conclusion}

In this study, we investigated the effectiveness of various molecular feature representations for predicting Caco-2 permeability using AutoML-based modeling. Our key contributions are summarized as follows.

\textit{Comprehensive benchmarking of molecular representations}

We conducted a systematic evaluation of eight molecular feature representation types, including 2D/3D descriptors and fingerprints, across two datasets (TDC and OCHEM). This benchmarking revealed that PaDEL, Mordred, and RDKit descriptors are among the most effective for Caco-2 prediction, offering a practical reference for future model development. Detailed usage instructions for the CaliciBoost module on the Pharmaco-Net platform are provided in Supplementary Information 2.

\textit{Evidence of the importance of 3D descriptors}

Our experiments demonstrate that 3D descriptors contribute significantly to model performance. In both PaDEL and Mordred feature sets, models trained with combined 2D and 3D descriptors consistently outperformed those using only 2D features. Top features selected through SHAP and permutation importance analyses further confirmed the relevance of 3D components.

\textit{Development of a state-of-the-art model}

Leveraging the insights from our feature evaluation and AutoML optimization, we developed CaliciBoost model with a MAE of 0.2560, outperforming all previously reported methods.
These findings suggest that combining top-ranked molecular descriptors with AutoML-based modeling can provide robust and generalizable models for permeability prediction. For future work, combining multiple feature representation types - especially those shown to contribute meaningfully such as PaDEL, Mordred, and RDKit - may further enhance model performance. In addition, expanding the dataset by incorporating Caco-2 measurements from alternative sources like PubChem BioAssay could support training of more complex architectures, including deep graph-based models, and help address current limitations in data diversity and volume.

\section{Availability of data and materials} \label{sec:style}

The TDC dataset used in this study is publicly available from the Therapeutics Data Commons (TDC) Caco-2 task page: 
\url{https://tdcommons.ai/single_pred_tasks/adme/#caco-2-cell-effective-permeability-wang-et-al}

The OCHEM dataset was curated, filtered, and processed by the authors, and is available at: 
\url{https://huggingface.co/datasets/junhong1222/Caco2-Ochem-dataset}

The GitHub repository containing the full implementation, including code and pretrained model for CaliciBoost is accessible at: 
\url{https://github.com/Calici/CaliciBoost}

%% Please use the acknowledgment and contribution environments. This will 
%% be anonomyized when the "anonymous" style option is used. 
\section{Acknowledgments}
H.V.L. gratefully acknowledges the valuable advice and encouragement provided by Dr. Quoc-Khanh Nguyen throughout the development of this research work.

\section{Funding}
This research was supported by the Ministry of Food and Drug Safety (MFDS), Republic of Korea, through a grant (Project No. RS-2024-00332490) and the Graduate School Education Program of Regulatory Sciences for Functional Food (21153MFDS604) in 2025. It was also supported by a grant of the Korea Health Technology R\&D Project through the Korea Health Industry Development Institute (KHIDI), funded by the Ministry of Health \& Welfare, Republic of Korea (Grant No. RS-2022-KH130308). Additional financial support was provided by the Ministry of Trade, Industry, and Energy (MOTIE), Korea, under the “Infrastructure Program for Industrial Innovation” supervised by the Korea Institute for Advancement of Technology (KIAT). Furthermore, this study was supported by the 2025 project titled “Establishing the Foundation for In Silico Industrialization in Uiseong-gun, Gyeongsangbuk-do.”

\bibliography{sample701}{}

\begin{thebibliography}{}
\expandafter\ifx\csname natexlab\endcsname\relax\def\natexlab#1{#1}\fi
\providecommand{\url}[1]{\href{#1}{#1}}
\providecommand{\dodoi}[1]{doi:~\href{http://doi.org/#1}{\nolinkurl{#1}}}
\providecommand{\doeprint}[1]{\href{http://ascl.net/#1}{\nolinkurl{http://ascl.net/#1}}}
\providecommand{\doarXiv}[1]{\href{https://arxiv.org/abs/#1}{\nolinkurl{https://arxiv.org/abs/#1}}}

% type= article
\bibitem[{L. Breiman(2001)Breiman}]{ref27}
Breiman, L. 2001, \bibinfo{title}{Random forests,} Machine learning, 45, 5

% type= article
\bibitem[{N. Erickson {et~al.}(2020)Erickson, Mueller, Shirkov, Zhang, Larroy, Li, \& Smola}]{ref20}
Erickson, N., Mueller, J., Shirkov, A., {et~al.} 2020, \bibinfo{title}{Autogluon-tabular: Robust and accurate automl for structured data,} arXiv preprint arXiv:2003.06505

% type= article
\bibitem[{M. Feurer {et~al.}(2022)Feurer, Eggensperger, Falkner, Lindauer, \& Hutter}]{ref16}
Feurer, M., Eggensperger, K., Falkner, S., Lindauer, M., \& Hutter, F. 2022, \bibinfo{title}{Auto-sklearn 2.0: Hands-free automl via meta-learning,} Journal of Machine Learning Research, 23, 1

% type= article
\bibitem[{P. Gedeck {et~al.}(2006)Gedeck, Rohde, \& Bartels}]{ref7}
Gedeck, P., Rohde, B., \& Bartels, C. 2006, \bibinfo{title}{QSAR- how good is it in practice? Comparison of descriptor sets on an unbiased cross section of corporate data sets,} Journal of chemical information and modeling, 46, 1924

% type= article
\bibitem[{Y. Gui {et~al.}(2024)Gui, Zhan, \& Li}]{ref21}
Gui, Y., Zhan, D., \& Li, T. 2024, \bibinfo{title}{Taking another step: A simple approach to high-dimensional Bayesian optimization,} Information Sciences, 679, 121056

% type= article
\bibitem[{H. Hadipour {et~al.}(2022)Hadipour, Liu, Davis, Cardona, \& Hu}]{ref10}
Hadipour, H., Liu, C., Davis, R., Cardona, S.~T., \& Hu, P. 2022, \bibinfo{title}{Deep clustering of small molecules at large-scale via variational autoencoder embedding and K-means,} BMC bioinformatics, 23, 132

% type= article
\bibitem[{H. Kuwahara \& X. Gao(2021)Kuwahara \& Gao}]{ref12}
Kuwahara, H., \& Gao, X. 2021, \bibinfo{title}{Analysis of the effects of related fingerprints on molecular similarity using an eigenvalue entropy approach,} Journal of cheminformatics, 13, 1

% type= inproceedings
\bibitem[{E. LeDell \& S. Poirier(2020)LeDell \& Poirier}]{ref18}
LeDell, E., \& Poirier, S. 2020, \bibinfo{title}{H2o automl: Scalable automatic machine learning,} in Proceedings of the AutoML Workshop at ICML, Vol. 2020, 24

% type= article
\bibitem[{S.~M. Lundberg \& S.-I. Lee(2017)Lundberg \& Lee}]{ref28}
Lundberg, S.~M., \& Lee, S.-I. 2017, \bibinfo{title}{A unified approach to interpreting model predictions,} Advances in neural information processing systems, 30

% type= inproceedings
\bibitem[{M. Malu {et~al.}(2021)Malu, Dasarathy, \& Spanias}]{ref22}
Malu, M., Dasarathy, G., \& Spanias, A. 2021, \bibinfo{title}{Bayesian optimization in high-dimensional spaces: A brief survey,} in 2021 12th International Conference on Information, Intelligence, Systems \& Applications (IISA), IEEE, 1--8

% type= article
\bibitem[{H.~L. Morgan(1965)Morgan}]{ref6}
Morgan, H.~L. 1965, \bibinfo{title}{The generation of a unique machine description for chemical structures-a technique developed at chemical abstracts service.,} Journal of chemical documentation, 5, 107

% type= article
\bibitem[{H. Moriwaki {et~al.}(2018)Moriwaki, Tian, Kawashita, \& Takagi}]{ref3}
Moriwaki, H., Tian, Y.-S., Kawashita, N., \& Takagi, T. 2018, \bibinfo{title}{Mordred: a molecular descriptor calculator,} Journal of cheminformatics, 10, 4

% type= inproceedings
\bibitem[{R.~S. Olson \& J.~H. Moore(2016)Olson \& Moore}]{ref17}
Olson, R.~S., \& Moore, J.~H. 2016, \bibinfo{title}{TPOT: A tree-based pipeline optimization tool for automating machine learning,} in Workshop on automatic machine learning, PMLR, 66--74

% type= misc
\bibitem[{ rdkit(2025)rdkit}]{ref4}
rdkit. 2025, rdkit, \url{https://www.rdkit.org/}

% type= article
\bibitem[{N. Stiefl {et~al.}(2006)Stiefl, Watson, Baumann, \& Zaliani}]{ref8}
Stiefl, N., Watson, I.~A., Baumann, K., \& Zaliani, A. 2006, \bibinfo{title}{ErG: 2D pharmacophore descriptions for scaffold hopping,} Journal of chemical information and modeling, 46, 208

% type= article
\bibitem[{I. Sushko {et~al.}(2011)Sushko, Novotarskyi, K{\"o}rner, Pandey, Rupp, Teetz, Brandmaier, Abdelaziz, Prokopenko, Tanchuk, {et~al.}}]{ref26}
Sushko, I., Novotarskyi, S., K{\"o}rner, R., {et~al.} 2011, \bibinfo{title}{Online chemical modeling environment (OCHEM): web platform for data storage, model development and publishing of chemical information,} Journal of computer-aided molecular design, 25, 533

% type= misc
\bibitem[{ TDC.Caco2\_Wang(2025)TDC.Caco2\_Wang}]{ref5}
TDC.Caco2\_Wang. 2025, TDC.Caco2\_Wang, \url{https://tdcommons.ai/benchmark/admet_group/01caco2/}

% type= article
\bibitem[{C. Wang {et~al.}(2021)Wang, Wu, Weimer, \& Zhu}]{ref19}
Wang, C., Wu, Q., Weimer, M., \& Zhu, E. 2021, \bibinfo{title}{Flaml: A fast and lightweight automl library,} Proceedings of Machine Learning and Systems, 3, 434

% type= article
\bibitem[{N.-N. Wang {et~al.}(2016)Wang, Dong, Deng, Zhu, Wen, Yao, Lu, Wang, \& Cao}]{ref25}
Wang, N.-N., Dong, J., Deng, Y.-H., {et~al.} 2016, \bibinfo{title}{ADME properties evaluation in drug discovery: prediction of Caco-2 cell permeability using a combination of NSGA-II and boosting,} Journal of chemical information and modeling, 56, 763

% type= article
\bibitem[{R. Winter {et~al.}(2019)Winter, Montanari, No{\'e}, \& Clevert}]{ref15}
Winter, R., Montanari, F., No{\'e}, F., \& Clevert, D.-A. 2019, \bibinfo{title}{Learning continuous and data-driven molecular descriptors by translating equivalent chemical representations,} Chemical science, 10, 1692

% type= article
\bibitem[{F. Wong {et~al.}(2024)Wong, Zheng, Valeri, Donghia, Anahtar, Omori, Li, Cubillos-Ruiz, Krishnan, Jin, {et~al.}}]{ref9}
Wong, F., Zheng, E.~J., Valeri, J.~A., {et~al.} 2024, \bibinfo{title}{Discovery of a structural class of antibiotics with explainable deep learning,} Nature, 626, 177

% type= article
\bibitem[{Z. Wu {et~al.}(2019)Wu, Lei, Shen, Wang, Cao, \& Hou}]{ref2}
Wu, Z., Lei, T., Shen, C., {et~al.} 2019, \bibinfo{title}{ADMET evaluation in drug discovery. 19. Reliable prediction of human cytochrome P450 inhibition using artificial intelligence approaches,} Journal of chemical information and modeling, 59, 4587

% type= article
\bibitem[{K. Yang {et~al.}(2019)Yang, Swanson, Jin, Coley, Eiden, Gao, Guzman-Perez, Hopper, Kelley, Mathea, {et~al.}}]{ref11}
Yang, K., Swanson, K., Jin, W., {et~al.} 2019, \bibinfo{title}{Analyzing learned molecular representations for property prediction,} Journal of chemical information and modeling, 59, 3370

% type= article
\bibitem[{C.~W. Yap(2011)Yap}]{ref13}
Yap, C.~W. 2011, \bibinfo{title}{PaDEL-descriptor: An open source software to calculate molecular descriptors and fingerprints,} Journal of computational chemistry, 32, 1466

% type= article
\bibitem[{S. Yee(1997)Yee}]{ref1}
Yee, S. 1997, \bibinfo{title}{In vitro permeability across Caco-2 cells (colonic) can predict in vivo (small intestinal) absorption in man—fact or myth,} Pharmaceutical research, 14, 763

\end{thebibliography}
\bibliographystyle{aasjournalv7}



\end{document}